# Detection of long-range orbital-Hall torques

Arnab Bose[1], Fabian Kammerbauer[1], Rahul Gupta[1], Dongwook Go[2], Yuriy Mokrousov[1,2], Gerhard Jakob[1,3], Mathias Kläui[1,3,4]

[1]Institute of Physics, Johannes Gutenberg University Mainz, Staudingerweg 7, 55128 Mainz, Germany

[2]Peter Grünberg Institut and Institute for Advanced Simulation, Forschungszentrum Jülich and JARA, 52425 Jülich, Germany

[3]Graduate School of Excellence Materials Science in Mainz, 55128 Mainz, Germany

[4]Center for Quantum Spintronics, Department of Physics, Norwegian University of Science and Technology, NO-7491 Trondheim, Norway

Abstract

*We report and quantify a large orbital-Hall torque generated by Nb and Ru, which we identify from the strong dependence of torques on the ferromagnets. This is manifested as a sign-reversal and strong enhancement in the damping-like torques measured in Nb (or Ru)/Ni bilayers as compared to Nb (or Ru)/FeCoB bilayers. The long-range nature of orbital transport in the ferromagnet is revealed by the thickness dependences of Ni in Nb (or Ru)/Ni bilayers which are markedly different from the regular spin absorption in the ferromagnet that takes place within few angstroms and thus it uniquely distinguishes the orbital Hall torque from the conventional spin Hall torque.*

The nonequilibrium flow of angular momentum has been one of the key aspects of condensed matter physics as it plays a major role in modern solid-state magnetic devices [1,2]. Nature provides two different types of intrinsic angular momenta in a material that can be relatively easily accessible for applications, which are: (1) orbital-momentum and (2) spin-momentum. In the past decade, the main focus of spintronics has been to inject spin-momenta into a magnet for non-volatile memory applications [1,2], which was triggered by the discovery of the spin-Hall effect (SHE) [3], a mechanism that generates a transverse spin current ($J_{SH}$) which can interact with the magnet directly via a spin-transfer torque (STT) [1]. However, this scheme of strong $J_{SH}$ generation is mostly limited to certain materials due to the requirement of large spin-orbit coupling (SOC) such as Pt, W etc [1]. Theory works have predicted the generation of large orbital-Hall current ($J_{OH}$) from orbital-Hall effect (OHE), not relying on the SOC of the non-magnetic material (NM) [4–13]. However, $J_{OH}$ remain challenging for an unambiguous experimental detection [14] since $J_{OH}$ does not interact directly with the magnetization of commonly studied ferromagnets (FM) [7–9], unlike $J_{SH}$ which generates STT in a magnetic layer that is largely independent of the FM [1]. So far, the orbital-Hall torques (OHT) have often been studied in systems with the naturally grown oxides such as $CuO_x$ and $AlO_x$ as the non-magnetic layer [15–19], and these oxide layers are often not well-controlled, making it difficult to compare the results with the theoretical calculations. So, it is a prime interest to study this effect in the clean and well-defined elements for a direct and unambiguous comparison.

While the spin-orbit torques (SOT) are essential ingredients for the memory application there have been strong disagreements in the predicted and experimentally measured torques [1] in some systems including the sign reversal [20–23], suggesting an important piece of physics is still missing. By the SHE mechanism longitudinal electric current ($J_C$) flowing along the $x$-direction in the heavy metals (HM) generates the flow of spins perpendicular to it (along the $z$-axis, sample growth direction) while the spins are polarized along $y$ (Fig. 1(a)). Similarly, in certain nonmagnets it has been predicted to have the flow of transverse orbital momenta (along the $z$-axis) from the $J_C$ (along the $x$-axis) with the orbital quantization axis being along $y$ [5] (Fig. 1(a), the orbital moment is normal to the drawn circles). This is referred to as orbital-Hall current ($J_{OH}$). Its advantage is that large values of $J_{OH}$ can be found in abundant material uncorrelated to the SOC that could be used for practical application.

The actions of $J_{SH}$ and $J_{OH}$ are distinctly different on the FM. $J_{SH}$ can directly interact with the static magnetization of the adjacent FM and thereby produces the damping-like torque (DLT) nearly independent of the FM in the commonly studied HM/FM bilayers due to the comparable magnitude of the spin-transparency [1]. On the other hand, $J_{OH}$ does not directly interact with the FM since the orbital momentum is quenched in the equilibrium state. However, it is recently predicted that the injected $J_{OH}$ can be converted into the spin-current ($J_{OH \to S}$) inside some of the FM using the SOC of the FM as schematically shown in Fig. 1(b) and hence $J_{OH \to S}$ would produce a torque on the FM, referred to as orbital-Hall torque (OHT) [7–9]. Ni is predicted to be very efficient for OHT while Fe is quite inefficient, suggesting a strong ferromagnet dependence as predicted in the recent theoretical [7–9] and reported in the previous experimental work [23] and also corroborated in this work by comparing NM/Ni and NM/Fe$_{60}$Co$_{20}$B$_{20}$ bilayers.

Another key difference between $J_{OH}$ and $J_{SH}$ is the length scale of the angular momenta (spin or/and orbital) transport inside the FM [9]. For example, the transverse component of the spins is absorbed within the first few monolayers of the FM [1,24,25] whereas the $J_{OH}$ is predicted to be transmitted over a long range inside the FM before it is fully converted into the spin-current ($J_{OH \to S}$) and hence expected to result in a long-range torque effect [9]. By systematically varying the thickness of Ni in NM/Ni bilayer we quantify the long-range torques which is an important outcome of this work and has not been reported in previously studied systems [15,18,23,26–28]. In presence of both $J_{OH}$ and $J_{SH}$ we can phenomenologically write the expression for the net damping-like torque ($\xi_{net}$) as follows:

$$\xi_{net} = \xi_{SH} + \xi_{OH}(1 - \text{sech}(t_{FM}/\lambda_{FM})) \qquad (1)$$

where $\xi_{SH}$ represents the conventional DLT as observed in the standard HM/FM bilayers in the absence of $J_{OH}$. $\xi_{OH}$ represents the DLT due to the $J_{OH \to S}$ which is strongly FM dependent [7–9]. The term $(1 - \text{sech}(t_{FM}/\lambda_{FM}))$ suggests the long-range action of the torques and uniquely distinguishes OHT from the regular SHT as evident in our experiment. $t_{FM}$ is the thickness of the FM while $\lambda_{FM}$ sets the length scale of the long-range torques. We find that $\lambda_{FM}$ of Ni is approximately 2.5 nm suggesting that it takes 8-10 nm Ni-thickness to get the full strength of the OHT, nearly an order of magnitude larger than the length scale of the conventional SHT [24,25].

While the long-range torque is the primary focus of this work, we have also explored the $J_{OH \to S}$ using the SOC of a HM (Pt in this case) rather than relying on the SOC of the FM by judiciously designing the stack: NM/Pt(t)/FM/Pt(t)/cap (Fig. 1(c)). Our proposed device design eliminates the contribution from the regular SHE due to the symmetric placement of Pt on both sides of the FM which was not considered in the previous reports [14,15,23,26,28] and thus our work presents a more comprehensive understanding of the fundamental interplay of $J_{OH}$ and $J_{SH}$ in spin-orbit coupled systems.

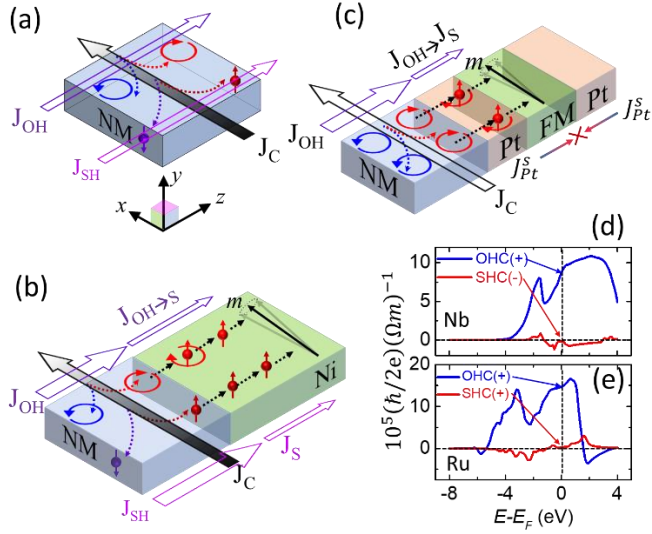

FIG. 1. (a) Schematics of the generation of $J_{OH}$ from the OHE and $J_{SH}$ from the SHE driven by $J_C$ in NM. (b) Two different components of spin-currents responsible for the SOT in the magnet: (1) $J_S$, the fraction of $J_{SH}$ injected into the FM and (2) $J_{OH\to S}$, the converted spin-current inside the FM from the generated $J_{OH}$. (c) Implementation of a NM/Pt/FM/Pt heterostructure. Externally injected $J_{OH}$ from the NM (low SOC) is converted into $J_{OH\to S}$ using the SOC of Pt that further produces a torque on FM. $J_{SH}$ produced from both sides of the Pt nearly cancel each other. Calculated orbital-Hall conductivity (OHC) and spin-Hall conductivity (SHC) of Nb (d) and Ru (e) by DFT. The OHC is large and positive for both Nb and Ru at Fermi level. The SHC is very small for both and negative for Nb and positive for Ru.

We study the NM, Nb and Ru as the source of $J_{OH}$ as they are predicted to generate a large orbital Hall conductivity (OHC) and negligible spin Hall conductivity (SHC) by density functional theory (DFT) (Fig. 1(d,e)) that allows for a clean experimental detection of OHE [4,6,29]. Four different sets of samples are prepared on the high resistive Si/SiO$_2$ wafer by the sputtering technique, set-1: Ru (or Nb)/Ni/cap, set-2: Ru (or Nb)/Fe$_{60}$Co$_{20}$B/cap, set-3: Ru (or Nb)/Pt($t_{Pt}$)/Fe$_{60}$Co$_{20}$B (or Ni$_{81}$Fe$_{19}$)/Pt($t_{Pt}$)/cap, set-4: Ru (or Nb)/Ni/Ru (or Nb)/cap. The thickness of Ru and Nb is 4 nm and 5 nm respectively. The thickness of Ni was varied from 4 to 12 nm. The thickness of Fe$_{60}$Co$_{20}$B$_{20}$ (FCB) and Ni$_{81}$Fe$_{19}$ (Py) was varied from 4-6 nm. More details about the sample preparation can be found in the supplementary information.

We performed spin-torque ferromagnetic resonance (ST-FMR) [30–32] to quantify the SOT. A radio frequency (rf) ($f_0$=7-12 GHz) current is passed through the device and a dc voltage is measured while sweeping an in-plane external magnetic field ($H$) at an angle $\phi$ with respect to $J_C$. The magnet responds to the radio frequency Oersted field ($H_{Oe}$) and the spin current ($J_S$) produced by the adjacent layers and oscillates. Homodyne mixture of the oscillatory magnetoresistance due to anisotropic magnetoresistance (AMR) and oscillatory $J_C$ produces a dc voltage [30–32] that is a superposition of symmetric ($V_S = S\left(\frac{\Delta^2}{(H-H_0)^2+\Delta^2}\right)$) and anti-symmetric ($V_A = A\left(\frac{(H-H_0)\Delta}{(H-H_0)^2+\Delta^2}\right)$) Lorentzian lines. $\Delta$ is the magnetic linewidth, $H_0$ is the resonant field, $S$, $A$ arise from the in-plane and out-of-plane torques, respectively. Unlike in single crystal materials, here we do not expect any extraordinary torques [33] as the sputtered films are typically polycrystalline. Hence the dominant origin of *S* is the in-plane-damping like torque (ip-DLT) generated from the injected $J_S$ into the FM whereas *A* originates from the out-of-plane field-like torques (op-FLT) generated by the $H_{Oe}$ and interfacial spin-orbit fields [32,34]. The efficiency of ST-FMR [32] can be defined as

$$\xi_{FMR} = \frac{S}{A} \frac{e\mu_0 M_S t_{NM} t_{FM}}{\hbar} \sqrt{1 + \left(\frac{M_{eff}}{H_0}\right)}, \qquad (2)$$

where $\mu_0$ is the permeability, $t_{NM}$ is the thickness of the NM, $M_S$ and $M_{eff}$ are the saturation magnetization and out-of-plane demagnetization field of the ferromagnet as determined by SQUID magnetometry and using Kittel's equation [30] (see Supplementary materials for a detailed experimental technique). In the absence of interfacial spin-orbit-fields (meaning $H_{Oe}$ is the dominant source of op-FLT) $\xi_{FMR}$ becomes the measure of the DLT efficiency per unit current density, $\xi_{DL,j}$ [30,32]. DLT efficiency per unit longitudinal field (equivalent to the effective spin-torque conductivity) can be calculated as: $\xi_{DL,E} = \xi_{DL,j}/\rho_{xx}$ where $\rho_{xx}$ is the longitudinal resistivity of the non-magnet. $\rho_{xx}$ of Nb (5 nm) and Ru (4 nm) are 41 μΩ-cm and 27.3 μΩ-cm, respectively, as calculated from the four-probe measurement technique.

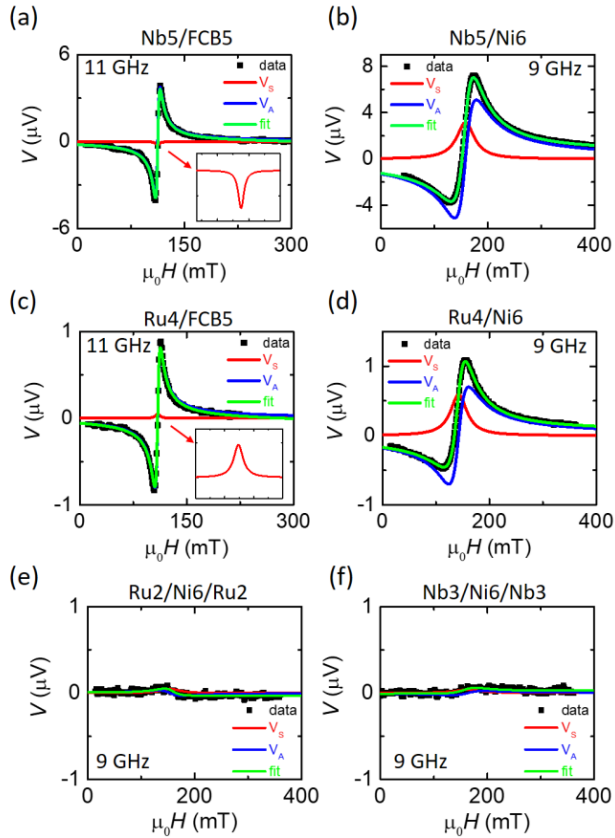

Fig. 2. Experimentally measured ST-FMR data in Nb(5 nm)/Fe$_{60}$Co$_{20}$B$_{20}$(FCB) (5 nm) (a), Nb (5 nm)/Ni (6 nm) (b), Ru (4 nm)/FCB (5 nm) (c), Ru (4 nm)/Ni (6 nm) (d), Ru (2 nm)/Ni (6 nm)/Ru (2 nm) (e) and Nb (3 nm)/Ni (6 nm)/Nb (3 nm) (f). Black points are the experimental data which is fit to the green curve that is the sum of $V_S$ (red curve), $V_A$ (blue curve) and a constant dc offset.

Fig. 2(a-d) shows the ST-FMR voltage spectra as a function of external field ($H$) sweep for Nb and Ru with the FM as FCB and Ni. Fig. 2(a,c) shows a small $V_S$ signal (red curves) estimating a negligible DLT ($|\xi_{DL}| \approx 0.001$) produced by both Nb and Ru. From the sign of $V_S$ (zoomed inset of Fig. 2(a,c)), we confirm that Nb exhibits a negative sign of SHE (same sign as W [35]) and Ru exhibits a positive sign (same sign as Pt [30]) but much smaller in magnitude which is consistent with the theoretical predictions [4,10] (Fig. 1(d,e)). When the Ni is used as a FM, we find a dramatic increase of $V_S$ for both Nb (Fig. 2(b)) and Ru (Fig. 2(d)) indicating a large DLT with positive sign for both cases that cannot be explained by the SHE. The enhanced DLT (for both Ru and Nb) including the sign change in $V_S$ (for Nb) is consistent with equation (1) suggesting a large $J_{OH}$ generated by the NM which is manifested as the OHT using Ni as a ferromagnetic detector while dominating over the SHT. Fig. 2(e,f) show the results when Ni is sandwiched between the symmetric layers (Ru/Ni/Ru and Nb/Ni/Nb samples). The measurements shows that the self-induced torques, resonant heating and other thermal artifacts are negligible [36]. Our data cannot be explained by the "self-induced torques" as that would lead to the opposite sign of the DLT as we observe for NM/Ni bilayers considering the sign of the SHE of Ni [37].

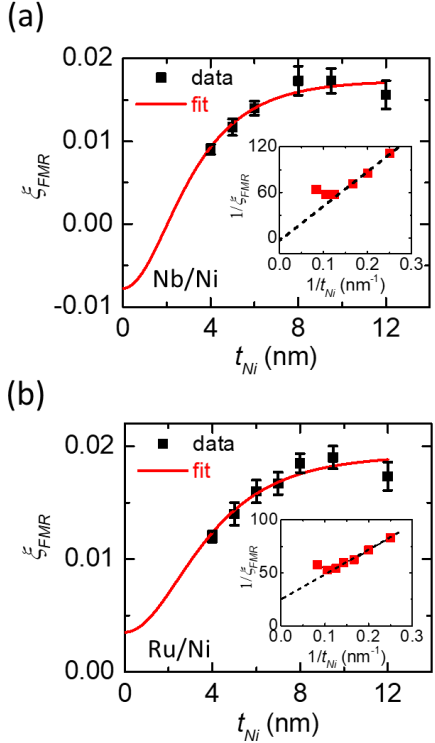

FIG. 3. $\xi_{FMR}$ as a function of the thickness of Ni ($t_{Ni}$) in Nb/Ni (a) and Ru/Ni (b) bilayers. The inset shows a plot of $1/\xi_{FMR}$ against $1/t_{Ni}$. The red continuous line is the fit using equation (1).

Fig. 3 shows the variations of $\xi_{FMR}$ as a function of Ni thickness ($t_{Ni}$) for both Nb (Fig. 3(a)) and Ru (Fig. 3(b)), which can be well described by equation (1). By fitting the experimental data with Eqn-1, we can quantify the contributions of SHE and OHE. The SHE contribution ($\xi_{SH}$) is estimated to be small for both Nb ($\xi_{SH} \approx -0.007$) and Ru ($\xi_{SH} \approx +0.003$), which is qualitatively consistent with Fig.

2(a,c) and DFT calculations (Fig. 1(d,e)) [4,10]. The estimated OHT efficiencies per unit current ($\xi_{OH,j}$) and field ($\xi_{OH,E}$) are approximately $\xi_{OH,J}^{Nb} \approx 0.025$ ($\xi_{OH,E}^{Nb} \approx \frac{\hbar}{2e} 6.1 \times 10^4\ (\Omega m)^{-1}$) and $\xi_{OH,J}^{Ru} \approx 0.016$ ($\xi_{OH,E}^{Ru} \approx \frac{\hbar}{2e} 5.86 \times 10^4\ (\Omega m)^{-1}$) for Nb and Ru respectively. From this fit, we calculate $\lambda_{Ni}$ of Ni is $2.2 \pm 0.7$ nm and $2.7 \pm 0.9$ nm in Nb/Ni and Ru/Ni samples, respectively suggesting that 8-10 nm Ni is needed to get the full strength of OHT from $J_{OH \to S}$ (Fig. 2). This length scale is substantially larger than for the SHT [24,25] and uniquely distinguishes OHT from the SHT. The inset of Fig. 2 shows plots of $1/\xi_{FMR}$ vs $1/t_{Ni}$ which has been a standard way to quantify the $\xi_{DL}$ using the ST-FMR measurements in presence of the interfacial FLT [32]. Using this technique, we find that $\xi_{DL}$ approaches to infinity for Nb/Ni which is nearly impossible highlighting the importance of the proposed alternate expression for the torque (Eqn-1). Note that in this analysis, we consider only the DLT produced by the effective spin current (Eqn. 1) and neglect interfacial op-FLT from the spin-current as it is generally small [34]. However, the proposed model (Eqn. 1) can fairly explain the thickness dependence of the measured torques (Fig. 3). We find that the AMR of Ni-based films is very large (more than 1%, largest among Py, Co, Fe and FeCoB) that enhances the rectified dc voltage from the spin-torques where spin-pumping effects would be negligible [36].

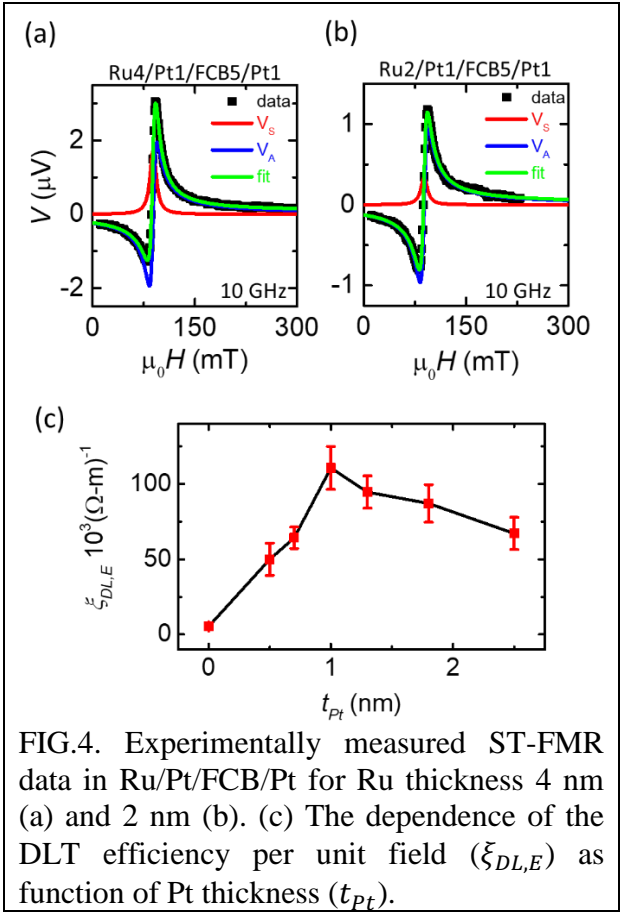

FIG.4. Experimentally measured ST-FMR data in Ru/Pt/FCB/Pt for Ru thickness 4 nm (a) and 2 nm (b). (c) The dependence of the DLT efficiency per unit field ($\xi_{DL,E}$) as function of Pt thickness ($t_{Pt}$).

In this section we show the conversion of $J_{OH \to S}$ using the SOC of a HM (Pt in this case) that can produce a torque on any FM for broader technological application. We implement devices as shown in Fig. 1(c) with Ru being the source of $J_{OH}$ and Pt being the converter of $J_{OH \to S}$. Even though Ru exhibits a negligible SHE (Fig. 2(c)) ($\xi_{DL,J} \approx +0.001$), a sizable DLT is observed in Ru/Pt(1)/FCB/Pt(1) heterostructures, $\xi_{DL,J} = 0.03 \pm 0.004$ for Ru thickness 4 nm (Fig. 4(a)) and $\xi_{DL,J} = 0.029 \pm 0.004$ for Ru thickness 2 nm (Fig. 4(b)), suggesting the $J_{OH}$ generation in the bulk of Ru by OHE. In this method the sign of DLT (hence $V_S$) should have the same sign of SHE of Pt since $J_{OH}$ is positive in these metals [4,10] which is also evident in Ru(4)/Pt(1)/FCB/Pt(1) samples (Fig. 4). Note that $J_{SH}$ generated by 1 nm Pt nearly cancels from both sides producing a negligible torques in Ru/Pt/FCB/Ru samples (see supplementary materials). Further we have quantified the $\xi_{DL,E}$ as a function of Pt-thickness (Fig. 4(c)) which shows the enhancement of DLT up to $t_{Pt}=1$ nm and then exponential decay of the DLT with the approximate estimated decay length of 2.5 nm. This behavior suggests that conversion of $J_{OH \to S}$ in the Pt taking place in much shorter length scale (< 1 nm) as compared to Ni (Fig. 3). For the Nb/Pt(1)/Py/Pt(1) samples we also observe a large enhancement of $V_S$ but with the opposite sign which cannot be explained by the OHE. We speculate this could be due to the interface generated spin [38,39] and/or orbital current [15,40]. While this is an interesting finding in itself, but not the main focus of this work and hence we provide more discussions in the supplementary materials.

| Net DLT | Nb/FeCoB | Ru/FeCoB | Nb/Ni | Ru/Ni | Ru/Pt/FeCoB/Pt |
|---|---|---|---|---|---|
| $\xi_{DL,J}$ | $-0.001 \pm 0.001$ | $+0.001 \pm 0.001$ | $+0.018 \pm 0.005$ | $+0.019 \pm 0.005$ | $+0.03 \pm 0.004$ |
| $\xi_{DL,E}\ 10^3 \frac{\hbar}{2e} (\Omega m)^{-1}$ | $-2.4 \pm 2.4$ | $+3.6 \pm 3.6$ | $+44 \pm 12$ | $+69.6 \pm 18.3$ | $+110 \pm 15$ |

Table-1. The efficiencies of DLT per unit current density ($\xi_{DL,J}$) and per unit electric field ($\xi_{DL,E}$) for different samples suggesting a negligible SHT in Nb (or Ru)/FeCoB samples, sizable DLT (SHT combined with OHT) in Nb (or Ru)/Ni samples (including the sign change in Nb) and large DLT in Ru/Pt/FeCoB/Pt samples driven by $J_{OH}$.

In summary, we report the generation of large $J_{OH}$ by elemental Ru and Nb determining the lower bound of the OHC to be $\frac{\hbar}{2e}(6.1 \pm 1,5) \times 10^4\ (\Omega m)^{-1}$ and $\frac{\hbar}{2e}(5.86 \pm 1.4) \times 10^4\ (\Omega m)^{-1}$ respectively which is an order of magnitude larger than the measured SHC providing a direct agreement with the theoretical predictions which was not possible to ascertain from the previous measurements done on the oxide based non-magnetic films due to the poorly defined structure. We demonstrate two key signatures of OHT that uniquely distinguish it from the STT, (1) strong dependence of the OHT on the adjacent ferromagnet (including the sign reversal in some cases such as Nb/Ni) and (2) the long-range action of the OHT inside the ferromagnet which we find to be approximately 8-10 nm for Ni films for the Nb/Ni and Ru/Ni bilayers. In contrast, the STT could take place even in the monolayer of a ferromagnet. We finally unambiguously demonstrate that an additional HM layer can generate an efficient torque by converting $J_{OH}$ to $J_S$ providing the necessary flexibility for the device applications.

NOTES

During the manuscript preparation we are aware of a recent and relevant work of orbital-torque generated Ti [41].

ACKNOWLEDGEMENT

D.G. and Y.M. thank the Jülich Supercomputing Centre for providing computational resources under project jiff40. This work was funded by the Deutsche Forschungsgemein- schaft (DFG). A.B., F.K., R.G., G.J, and M.K. thank Graduate School of Excellence Materials Science in Mainz (MAINZ, GSC266), Spin+X (A01, A11, B02) TRR 173–268565370 and Project No. 358671374. Horizon 2020 Framework Programme of the European Commission under FETOpen Grant Agreement No. 863155 (s-Nebula) and the European Research Council Grant Agreement No. 856538 (3D MAGiC) Research Council of Norway through its Centers of Excellence funding scheme, project number 262633 "QuSpin.". AB thanks Alexander von Humboldt foundation postdoctoral fellowship.REFERENCES

[1] A. Manchon, J. Železný, I. M. Miron, T. Jungwirth, J. Sinova, A. Thiaville, K. Garello, and P. Gambardella, *Current-Induced Spin-Orbit Torques in Ferromagnetic and Antiferromagnetic Systems*, Rev. Mod. Phys. **91**, 035004 (2019).

[2] B. Dieny, I. L. Prejbeanu, K. Garello, P. Gambardella, P. Freitas, R. Lehndorff, W. Raberg, U. Ebels, S. O. Demokritov, J. Akerman, A. Deac, P. Pirro, C. Adelmann, A. Anane, A. V. Chumak, A. Hirohata, S. Mangin, S. O. Valenzuela, M. C. Onbaşlı, M. D'Aquino, G. Prenat, G. Finocchio, L. Lopez-Diaz, R. Chantrell, O. Chubykalo-Fesenko, and P. Bortolotti, *Opportunities and Challenges for Spintronics in the Microelectronics Industry*, Nat. Electron. **3**, 446 (2020).

[3] J. Sinova, S. O. Valenzuela, J. Wunderlich, C. H. Back, and T. Jungwirth, *Spin Hall Effects*, Rev. Mod. Phys. **87**, 1213 (2015).

[4] T. Tanaka, H. Kontani, M. Naito, T. Naito, D. S. Hirashima, K. Yamada, and J. Inoue, *Intrinsic Spin Hall Effect and Orbital Hall Effect in 4d and 5d Transition Metals*, Phys. Rev. B **77**, 165117 (2008).

[5] D. Go, D. Jo, C. Kim, and H. Lee, *Intrinsic Spin and Orbital Hall Effects from Orbital Texture*, Phys. Rev. Lett. **121**, 86602 (2018).

[6] D. Jo, D. Go, and H.-W. Lee, *Gigantic Intrinsic Orbital Hall Effects in Weakly Spin-Orbit Coupled Metals*, Phys. Rev. B **98**, 214405 (2018).

[7] D. Go and H.-W. Lee, *Orbital Torque: Torque Generation by Orbital Current Injection*, Phys. Rev. Res. **2**, 013177 (2020).

[8] D. Go, F. Freimuth, J.-P. Hanke, F. Xue, O. Gomonay, K.-J. Lee, S. Blügel, P. M. Haney, H.-W. Lee, and Y. Mokrousov, *Theory of Current-Induced Angular Momentum Transfer Dynamics in Spin-Orbit Coupled Systems*, Phys. Rev. Res. **2**, 033401 (2020).

[9] D. Go, D. Jo, K.-W. Kim, S. Lee, M.-G. Kang, B.-G. Park, S. Blügel, H.-W. Lee, and Y. Mokrousov, *Long-Range Orbital Magnetoelectric Torque in Ferromagnets*, ArXiv **2106.07928**, 1 (2021).

[10] L. Salemi and P. M. Oppeneer, *First-Principles Theory of Intrinsic Spin and Orbital Hall and Nernst Effects in Metallic Monoatomic Crystals*, Phys. Rev. Mater. **6**, 095001 (2022).

[11] S. Bhowal and S. Satpathy, *Intrinsic Orbital Moment and Prediction of a Large Orbital Hall Effect in Two-Dimensional Transition Metal Dichalcogenides*, Phys. Rev. B **101**, 1211128

(2020).

[12] L. M. Canonico, T. P. Cysne, A. Molina-Sanchez, R. B. Muniz, and T. G. Rappoport, *Orbital Hall Insulating Phase in Transition Metal Dichalcogenide Monolayers*, Phys. Rev. B **101**, 161409 (2020).

[13] A. Pezo, D. García Ovalle, and A. Manchon, *Orbital Hall Effect in Crystals: Interatomic versus Intra-Atomic Contributions*, Phys. Rev. B **106**, 104414 (2022).

[14] D. Go, D. Jo, H.-W. Lee, M. Kläui, and Y. Mokrousov, *Orbitronics: Orbital Currents in Solids*, EPL (Europhysics Lett. **135**, 37001 (2021).

[15] S. Ding, A. Ross, D. Go, L. Baldrati, Z. Ren, F. Freimuth, S. Becker, F. Kammerbauer, J. Yang, G. Jakob, Y. Mokrousov, and M. Kläui, *Harnessing Orbital-to-Spin Conversion of Interfacial Orbital Currents for Efficient Spin-Orbit Torques*, Phys. Rev. Lett. **125**, 177201 (2020).

[16] H. An, Y. Kageyama, Y. Kanno, N. Enishi, and K. Ando, *Spin–Torque Generator Engineered by Natural Oxidation of Cu*, Nat. Commun. **7**, 13069 (2016).

[17] T. Gao, A. Qaiumzadeh, H. An, A. Musha, Y. Kageyama, J. Shi, and K. Ando, *Intrinsic Spin-Orbit Torque Arising from the Berry Curvature in a Metallic-Magnet / Cu-Oxide Interface*, Phys. Rev. Lett. **121**, 17202 (2018).

[18] J. Kim, D. Go, H. Tsai, D. Jo, K. Kondou, H.-W. Lee, and Y. Otani, *Nontrivial Torque Generation by Orbital Angular Momentum Injection in Ferromagnetic-Metal/Cu/Al2O3 Trilayers*, Phys. Rev. B **103**, L020407 (2021).

[19] L. Liao, F. Xue, L. Han, J. Kim, R. Zhang, L. Li, J. Liu, X. Kou, C. Song, F. Pan, and Y. Otani, *Efficient Orbital Torque in Polycrystalline Ferromagnetic/Ru/Al$_2$O$_3$: Theory and Experiment*, Phys. Rev. B **105**, 104434 (2022).

[20] K. Ueda, C. F. Pai, A. J. Tan, M. Mann, and G. S. D. Beach, *Effect of Rare Earth Metal on the Spin-Orbit Torque in Magnetic Heterostructures*, Appl. Phys. Lett. **108**, 232405 (2016).

[21] A. Bose, H. Singh, V. K. Kushwaha, S. Bhuktare, S. Dutta, and A. A. Tulapurkar, *Sign Reversal of Fieldlike Spin-Orbit Torque in an Ultrathin Cr/Ni Bilayer*, Phys. Rev. Appl. **9**, 014022 (2018).

[22] A. Bose, J. N. Nelson, X. S. Zhang, P. Jadaun, R. Jain, D. G. Schlom, D. C. Ralph, D. A. Muller, K. M. Shen, and R. A. Buhrman, *Effects of Anisotropic Strain on Spin-Orbit Torque Produced by the Dirac Nodal Line Semimetal IrO2*, ACS Appl. Mater. Interfaces **12**, 55411 (2020).

[23] D. Lee, D. Go, H.-J. Park, W. Jeong, H.-W. Ko, D. Yun, D. Jo, S. Lee, G. Go, J. H. Oh, K.-J. Kim, B.-G. Park, B.-C. Min, H. C. Koo, H.-W. Lee, O. Lee, and K.-J. Lee, *Orbital Torque in Magnetic Bilayers*, Nat. Commun. **12**, 6710 (2021).

[24] D. C. Ralph and M. D. Stiles, *Spin Transfer Torques*, J. Magn. Magn. Mater. **320**, 1190 (2008).

[25] A. Brataas, A. D. Kent, and H. Ohno, *Current-Induced Torques in Magnetic Materials*, Nat. Mater. **11**, 372 (2012).

[26] S. Lee, M.-G. Kang, D. Go, D. Kim, J.-H. Kang, T. Lee, G.-H. Lee, J. Kang, N. J. Lee, Y. Mokrousov, S. Kim, K.-J. Kim, K.-J. Lee, and B.-G. Park, *Efficient Conversion of Orbital Hall Current to Spin Current for Spin-Orbit Torque Switching*, Commun. Phys. **4**, 234 (2021).

[27] Y. Choi, D. Jo, K. Ko, D. Go, and H. Lee, *Observation of the Orbital Hall Effect in a Light Metal Ti*, ArXiv:2109.14847v1 1 (2021).

[28] G. Sala and P. Gambardella, *Giant Orbital Hall Effect and Orbital-to-Spin Conversion in 3d, 5d, and 4f Metallic Heterostructures*, Phys. Rev. Res. **4**, 033037 (2022).

[29] L. Salemi and P. M. Oppeneer, *First-Principles Theory of Intrinsic Spin and Orbital Hall and Nernst Effects in Metallic Monoatomic Crystals*, ArXiv: **2203.17037**, 1 (2022).

[30] L. Liu, T. Moriyama, D. C. Ralph, and R. A. Buhrman, *Spin-Torque Ferromagnetic Resonance Induced by the Spin Hall Effect*, Phys. Rev. Lett. **106**, 036601 (2011).

[31] D. Fang, H. Kurebayashi, J. Wunderlich, K. Výborný, L. P. Zârbo, R. P. Campion, A. Casiraghi, B. L. Gallagher, T. Jungwirth, and A. J. Ferguson, *Spin-Orbit-Driven Ferromagnetic Resonance*, Nat. Nanotechnol. **6**, 413 (2011).

[32] C.-F. Pai, Y. Ou, L. H. Vilela-Leão, D. C. Ralph, and R. A. Buhrman, *Dependence of the Efficiency of Spin Hall Torque on the Transparency of Pt/Ferromagnetic Layer Interfaces*, Phys. Rev. B **92**, 064426 (2015).

[33] A. Bose, N. J. Schreiber, R. Jain, D.-F. Shao, H. P. Nair, J. Sun, X. S. Zhang, D. A. Muller, E. Y. Tsymbal, D. G. Schlom, and D. C. Ralph, *Tilted Spin Current Generated by the Collinear Antiferromagnet Ruthenium Dioxide*, Nat. Electron. **5**, 267 (2022).

[34] Y.-T. Chen, S. Takahashi, H. Nakayama, M. Althammer, S. T. B. Goennenwein, E. Saitoh, and G. E. W. Bauer, *Theory of Spin Hall Magnetoresistance*, Phys. Rev. B **87**, 144411 (2013).

[35] C.-F. Pai, L. Liu, Y. Li, H. W. Tseng, D. C. Ralph, and R. A. Buhrman, *Spin Transfer Torque Devices Utilizing the Giant Spin Hall Effect of Tungsten*, Appl. Phys. Lett. **101**, 122404 (2012).

[36] S. Karimeddiny, J. A. Mittelstaedt, R. A. Buhrman, and D. C. Ralph, *Transverse and Longitudinal Spin-Torque Ferromagnetic Resonance for Improved Measurement of Spin-Orbit Torque*, Phys. Rev. Appl. **14**, 024024 (2020).

[37] H. Wang, C. Du, P. Chris Hammel, and F. Yang, *Spin Current and Inverse Spin Hall Effect in Ferromagnetic Metals Probed by $Y_3Fe_5O_{12}$-Based Spin Pumping*, Appl. Phys. Lett. **104**, 7 (2014).

[38] V. P. Amin, J. Zemen, and M. D. Stiles, *Interface-Generated Spin Currents*, Phys. Rev. Lett. **121**, 136805 (2018).

[39] A. Manchon, H. C. Koo, J. Nitta, S. M. Frolov, and R. A. Duine, *New Perspectives for Rashba Spin-Orbit Coupling*, Nat. Mater. **14**, 871 (2015).

[40] D. Go, D. Jo, T. Gao, K. Ando, S. Blügel, H.-W. Lee, and Y. Mokrousov, *Orbital Rashba Effect in a Surface-Oxidized Cu Film*, Phys. Rev. B **103**, L121113 (2021).

[41] H. Hayashi, D. Jo, D. Go, Y. Mokrousov, H.-W. Lee, and K. Ando, *Observation of Long-Range Orbital Transport and Giant Orbital Torque*, ArXiv **2202.13896**, 1 (2022).